# `modelSolver`: A Symbolic Model-Driven Solver for Power Network Simulation and Monitoring

Izudin Džafić, *Senior Member, IEEE* and Rabih A. Jabr, *Fellow, IEEE*

*Abstract*—The development of advanced software tools for power system analysis requires extensive programming expertise. Even when using open-source tools, programming skills are essential to modify built-in models. This can be particularly challenging for domain experts who lack coding proficiency. This paper introduces `modelSolver`, a software solution with a new framework centered around symbolic mathematical modeling. The proposed paradigm facilitates defining models through intuitive mathematical expressions, thus eliminating the need for traditional programming constructs such as arrays, loops, and sparse matrix computations. The `modelSolver` focuses on power flow and state estimation using an open-box approach, which allows users to specify custom models using either real or complex variables. Unlike existing tools that rely on hard-coded models, modelSolver enables the representation of a wide range of advanced functionalities, including power flow with voltage regulators and load tap changers, continuation power flow, and Gauss-Newton state estimation with equality constraints. Compatibility with MATPOWER is ensured via a converter that automates importing data files. The framework prioritizes model-driven development and empowers domain experts to focus on power system modeling without programming barriers. It aims to simplify power system computations, making them more accessible to students, scientists, and practitioners.

## I. Introduction

Power system steady-state computing applications include essential tasks for power network solution and monitoring, such as power flow and state estimation. The computing tasks depend on specialized software tools created by commercial power system application vendors or commonly used open-source tools for power system education and research, such as PSAT [1] and MATPOWER [2]. The software tools encode specific models with predefined configurations, and updating these models requires familiarity with programming constructs such as loops, arrays, and sparse matrix operations, including constructing and factoring the sparse Jacobian matrix [3]. There is a growing interest in incorporating power flow and voltage control devices [4], [5] along with Inverter-Based Resources (IBRs) [6], [7] into network simulators. The role of IBRs in managing reactive power has been addressed in recent standards, including IEEE and national interconnection standards [8]–[10]. Research efforts have been made to incorporate specific model inverters in power flow applications for both grid-connected and islanded operational modes [11]. Volt-VAr curves have been designed specifically for use in particular grid-connected power flow solvers, including PowerWorld [12], [13] and other tailor-made power flow extensions [14]. Other efforts focused on modeling droop-controlled microgrids [15], [16], covering both AC and DC interconnections [17], [18], as well as unbalanced operations [19]. The models discussed above may not be included in existing software packages employed by power system engineers, which can create a barrier for domain experts. Although domain specialists have expertise in power system modeling, they often lack the programming skills necessary to modify and expand the capabilities of open-source software tools. They, therefore, rely on major software upgrades or adopt other tools to accomplish specific simulation tasks, which can be both inefficient and impractical. The situation is exacerbated with commercial tools: since their source code is proprietary, users are unable to modify the internal models. To meet diverse needs, industrial software suites often incorporate excessive features, resulting in outsized, complex, and costly solutions. This highlights a critical gap for researchers and engineers, primarily the need for a flexible modeling tool that does not necessitate extensive programming knowledge to customize models to specific applications and controller models. The need is particularly important with the proliferation of IBRs, as their inappropriate response to voltage control has been linked to recent system blackouts [20].

This manuscript addresses the above challenges by presenting `modelSolver`, a software featuring a new framework that allows domain experts to perform steady-state power system calculations, including power flow and state estimation. The computing problems can include arbitrary control models defined in symbolic form using mathematical expressions, thereby eliminating the need for traditional programming. The `modelSolver` is fundamentally different than closed-box systems where the models are hard-coded. It instead adopts an open-box philosophy, enabling domain experts to specify custom models tailored to distinct system requirements or new controller paradigms. The `modelSolver` is based on a declarative framework to define parameters, variables, and nonlinear algebraic equations (`NLEs`) in a natural symbolic form. The software then automatically carries out all computation steps, including forming and factorizing the sparse Jacobian matrix, irrespective of the models used in the `NLEs`. This functionality contrasts with popular open-source software packages used in research and teaching, such as MATPOWER [2], where implementing advanced controller models in power flow is not straightforward [3]. The overall approach is model-driven, enabling users to focus on power system domain

I. Džafić is with the University of Sarajevo, Faculty of Electrical Engineering, Zmaja od Bosne bb 71000 Sarajevo (email: idzafic@ieee.org).

R. A. Jabr is with the Department of Electrical & Computer Engineering, American University of Beirut, P.O. Box 11-0236, Riad El- Solh / Beirut 1107 2020, Lebanon (email: rabih.jabr@aub.edu.lb).

knowledge rather than implementation details.

The `modelSolver` offers some unique features that are not supported by existing open-source packages. First, it allows symbolic model definitions and subsequent solutions in either real or complex variables. Second, it supports statistical distributions that model measurement noise and a `SubModel` for simulating true measurements from power flow, as well as more complex computational features, such as equality-constrained state estimation and continuation power flow. The following significant contributions are provided by this work:

1) Open and symbolic modeling: without depending on hard-coded implementations, users can define custom models using natural syntax.
2) No coding required: domain experts focus on the model equations instead of programming constructs.
3) Model extensibility: new component and control models can be simulated without modifying the solver's internal routines.
4) MATPOWER compatibility: existing m-file power flow models can be imported via a converter in one of three possible formats (polar, rectangular, or complex).
5) Educational benefits: students learn power system modeling concepts without worrying about software code.

The rest of this paper is organized as follows. Section II offers an overview of the `modelSolver`. It includes a flowchart of the software framework along with explanations of hidden loops, internal variables, and points for reporting results. Section III illustrates power flow modeling using both real and complex variables. It demonstrates the use of two conditional statements, `if-else` and `switch-case`, through examples. The `if-else` statement is employed to represent PV nodes with reactive power limits. The `switch-case` statement is utilized to model piecewise Volt-VAr curves, which represent the contribution of IBRs to reactive power control. Power flow extensions are also discussed for iterative simulation, such as continuation power flow applications. Section IV presents state estimation modeling, including the setting of weights for the Weighted Least-Squares (`WLS`) solution, Equality Constraints (`ECs`) for modeling virtual measurements, and a `SubModel` for generating true measurements and modifying them with noise. MATPOWER converters and their applications are discussed in Section V. Section VI concludes the paper and provides a direction for future work aimed at enhancing the capabilities of the proposed solver.

## II. MODEL-DRIVEN POWER SYSTEM DEVELOPMENT

The proposed framework is built around a novel modeling syntax that simplifies power system modeling and analysis. It allows users to define models using a group- and attribute-oriented language. Each group in the syntax begins with a label that organizes model parts based on a common logical framework. There are three groups to start with: `Vars`, `Params`, `NLEs`/`WLSEs`.

`Vars`     group lists all variables to be solved in addition to their initial values (if specified).

`Params`     group lists all the parameters used in the model together with their values.

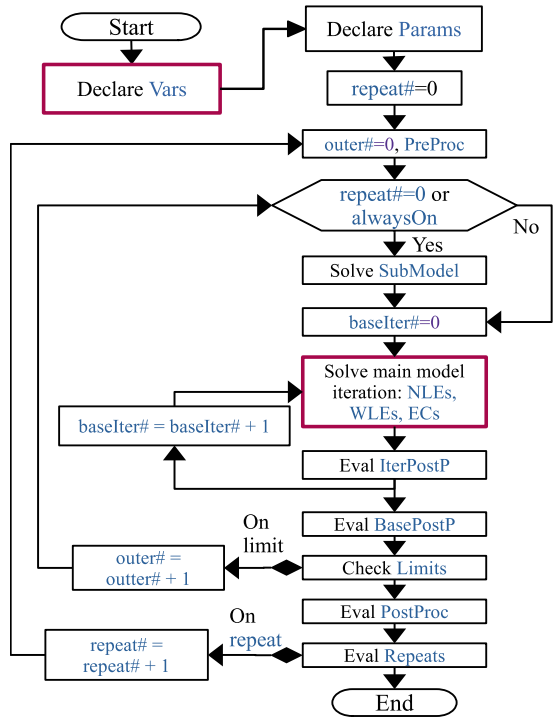

Fig. 1: The internal solution flow within the `modelSolver` framework (the blue elements represent group names and embedded variables, while the mandatory components are outlined with a red border).

`NLEs`     group for power flow defines the nonlinear algebraic equations in symbolic form, and it may include `if-else` and `switch-case` conditions to capture the model behavior.

`WLSEs`     group for state estimation defines the nonlinear algebraic equations of the measurement model together with the weights employed in `WLS` estimation.

Group names are reserved keywords and they cannot be used as variable or parameter names. The behavior of each group is adjusted through its attributes, which are placed in brackets. Each attribute must be initialized with a value, and if the attribute is repeated, the program considers only the first occurrence. Attributes appear on the same line as the group declaration and must be placed before the terminating token of the group (:). Attribute names may be reused for variable and parameter names. Fig. 1 shows the internal logic for the framework of the software, referred to as `modelSolver`, where the only mandatory components are those outlined with a red border. The framework employs three types of expressions.

1) Equation expressions are employed for defining the model. The use of the assignment operator (=) in equation expressions is optional. It can be included only once to equate the expression to a value, and when omitted, the equation expression is equal to zero. Equation expressions can only be included in the groups of `NLEs`, `WLSEs`, and `ECs`. The equations are parsed, and depending on the model, either a symbolic sparse Jacobian will be calculated and passed to the Newton-Raphson solver [21], or a symbolic sparse measurement Jacobian will be generated and passed to the sparse Gain matrix composer [22] for state estimation. The state estimation includes the normal equations approach with equality constraints, if applicable.

2) Assignment expressions are used to calculate and assign pre-, inter-, and post-processing values. These expressions appear in the groups for `BasePostP`, `IterPostP`, `Limits`, `PostProc`, `PreProc`, and `Repeats`. The first operand in assignment expressions can be a variable or a parameter. Unlike equation expressions, assignment expressions support several assignment operators: `=`, `+=`, `-=`, `*=`, `/=`, and `^=`. The syntax of these operators follows C/C++ assignment operator conventions.

3) Logic expressions are used for implementing two types of conditional statements (`if-else` and `switch-case`) for logical branching.

The reserved keywords in the `modelSolver` are listed below and also include embedded real and complex functions that are used to invoke standard mathematical functions, with their syntax similar to MATLAB.

`if-else`: for specifying logical branching statements.
`switch-case`: similar logic as `if-else`, but more compact and easier to read.
`default`: defines the `default` case in a `switch-case` expression. It must be the last case.
`group`: used to group limits in the `Limits` section.
`end`: used to terminate a model, group, or `if-else`/`switch-case` block.
`repeat`: used to trigger solution repetition as shown in Fig. 1.

The `modelSolver` is available for download as a free standalone software from [23]. The GitHub repository [23] contains accompanying test cases, including all those discussed in this paper.

## III. POWER FLOW MODELING AND EXTENSIONS

This section demonstrates the use of the `modelSolver` for power flow applications. It shows the modeling using either real or complex variables, including the use of conditional statements, e.g., for handling PV nodes with reactive power limits and Volt-VAr control in an IBR. Iterative simulations are also demonstrated for computing the loadability characteristic (P-V curve) through the `Repeats` group.

### A. Real and Complex Variables

*1) real variables:* Fig. 2 shows a three-node network including a slack node, a zero-injection node, and a PQ node. The unknown voltages are at nodes 2 and 3. Fig. 3 shows the code that implements power flow in real variables. The `Model` group on line 5 shows that the problem type is `NL` (Non-Linear) and that the solution domain is `real` with a termination tolerance of 1E-6. The `Vars` group on line 6 has the attribute `out=true`, indicating that the variables defined on line 7 would feature in the output file. The initial values for the variables are optional. Still, given the solution using the Newton-Raphson method [21], they are recommended to be set based on the physical understanding of the problem. The initial angles are set to $\delta\_1$ and the initial voltage magnitudes to `v_1`, where $\delta\_1$ and `v_1` are parameters defined as part of the `Params` group on line 8. The other problem parameters on lines 10-13 are used in writing the power flow equations in the `NLEs` group on line 14. The `NLEs` include the sum of real and imaginary current set to zero in lines 16 and 17, and the real and reactive power injection equations in lines 19 and 20. The code contains comments marked with //, as seen in lines 15 and 18. The code is run using the following command in the terminal: `./modelSolver PathToModFile`.

*2) complex variables:* Fig. 4 shows a code that solves the same power flow problem but in the complex variable domain. Although less known in power system modeling, as introduced in textbooks [21], the solution in complex variables is more straightforward for defining the power flow equations [24]. In this case, the `Model` group on line 5 includes an attribute that specifies the variable domain as complex (`domain=cmplx`). Additionally, the `Vars` group on line 6 has `conj=true`, indicating that the variables `v_2` and `v_3` defined on line 7 also include their conjugates as part of the problem variables. The advantage of complex variable modeling becomes apparent when contrasting the code in Figs. 3 and 4. Specifically, this is evident in the specification of zero injection `NLEs` (lines 15-16 in Fig. 4 versus lines 16-17 in Fig. 3) and PQ modeling (lines 18-19 in Fig. 4 versus lines 19-20 in Fig. 3).

### B. Conditional Statements

*1) if-else statement:* Fig. 5 is a modification of the 3-node system in Fig. 2, where node 2 is of the PV type and has prespecified minimum and maximum reactive power limits. The code snippet in Fig. 6 shows that the `Params` group now includes parameters relevant to the generator connected at node 2 in addition to the boolean parameter `cGen2Reg`, which is initialized as `true`, meaning the generator is regulating the voltage at node 2 to its specified value `V2_sp`. The `NLEs` group defines the real power generation using complex variables in line 11. The `if-else` statement uses the status of `cGen2Reg` to indicate whether the generator regulates voltage (line 13) or generates reactive power at the specified value `Q2_inj` (line 15). The parameter `Q2_inj` is initialized to 0 on line 4. However, its value is recalculated

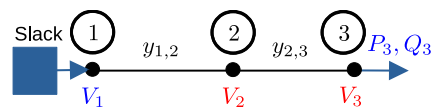

Fig. 2: A three-node network to demonstrate power flow modeling with a zero injection node and a PQ node (the blue text indicates known quantities, while the red text represents unknown variables).

```
1  Header:
2      maxIter=50
3      report=AllDetails //Solved - only final solved solution, All - shows solved and nonSolved
            with iterations, AllDetails - All + debug information
4  end  //end of header
5  Model [type=NL domain=real eps=1e-6 name="PF for 3 nodes (second node is Zero Injection)"]:
6  Vars [out=true]:
7      δ_2=δ_1; v_2=v_1; δ_3=δ_1; v_3=v_1 // Variable name and its initial value (optional)
8  Params:
9      δ_1=0; v_1=1 //slack angle and voltage (param)
10     aY=32.8797974610715; θ_diag=-1.40564764938027; θ_offDiag=1.73594500420952
11     aY11=aY; aY21=aY; aY22=2*aY; aY23=aY; aY32=aY; aY33=aY
12     θ_11=θ_diag; θ_21=θ_offDiag; θ_22=θ_diag; θ_23=θ_offDiag; θ_32=θ_offDiag; θ_33=θ_diag
13     P3_inj=-1; Q3_inj=-0.3
14 NLEs:
15     // node 2 (ZI). Instead of sum(powers) = 0 --> sum of currents = 0
16     aY21*v_1*cos(θ_21+δ_1) + v_2*aY22*cos(θ_22+δ_2) + aY23*v_3*cos(θ_23+δ_3)=0
17     aY21*v_1*sin(θ_21+δ_1) + v_2*aY22*sin(θ_22+δ_2) + aY23*v_3*sin(θ_23+δ_3)=0
18     // node 3: PQ, sum of powers = 0
19     v_3^2*aY33*cos(θ_33) + v_3*(aY32*v_2*cos(δ_3-θ_32-δ_2)) = P3_inj
20     -v_3^2*aY33*sin(θ_33) + v_3*(aY32*v_2*sin(δ_3-θ_32-δ_2)) = Q3_inj
21 end
```

Fig. 3: A code that demonstrates power flow modeling using real variables with a zero injection node and a PQ node (Example 1).

```
1  Header:
2      maxIter=50
3      report=AllDetails //Solved - only final solved solution, All - shows solved and nonSolved
            with iterations, AllDetails - All + debug information
4  end  //end of header
5  Model [type=NL domain=cmplx eps=1e-6 name="PF for 3 nodes (second node is Zero Injection)"]:
6  Vars [conj=true out=true]:
7      v2=v1; v3=v1
8  Params:
9      v1=1+0i
10     z12=0.005+0.03i; z23=0.005+0.03i;
11     y21=1/z12; y23=1/z23; y22=y21+y23; y33=y23
12     S3_inj=-1-0.3i
13 NLEs: //non linear equations
14     // node 2 (Sum of currents)
15     y22*v2-y21*v1-y23*v3=0       //sum of currents = 0
16     conj(y22*v2-y21*v1-y23*v3)=0  //sum of conj(currents) = 0
17     // node 3
18     v3*conj(y33*v3-y23*v2) = S3_inj
19     conj(v3)*(y33*v3-y23*v2) = conj(S3_inj)
20 end
```

Fig. 4: A code that demonstrates power flow modeling using complex variables with a zero injection node and a PQ node (Example 2).

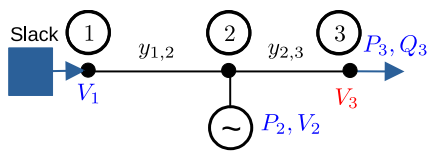

Fig. 5: A three-node network to demonstrate power flow modeling with a PV node and PQ node (the blue text indicates known quantities, while the red text represents unknown variables).

on line 24 within the `Limits` group. If any of the limits is violated, `Q2_inj` is then assigned either the minimum value `Q2_inj_min` (on line 27) or the maximum value `Q2_inj_max` (on line 31), and in this event, the `cGen2Reg` is set to `false`. When the generator's reactive power is set to one of its limits, the `modelSolver` re-executes the solution of the main `NLEs` model as shown in the framework of Fig. 1. The `group [name="Gen2"]` on line 21 is used to organize other conditional statements that are checked before the program returns to solving the `NLEs`. For instance, another generator can be included in the same group. It is also possible to create a separate group of conditions that could be checked

```
1  ...
2  Params:
3      ...
4      P2_inj=0.2; Q2_inj=0
5      V2_sp=1.01; Q2_inj_min=-1; Q2_inj_max=1.6 // Set to 1.6 to have voltage regulation
6      cGen2Reg=true [type=bool out=true]  // initially generator on node 2 is in regulation mode
7  // Start of outer loop
8  // Start of inner Loop
9  NLEs: //non linear equations
10     // node 2 (PV node): Sinj2 + conj(Sinj2) = 2*P2_inj
11     v2*conj(y22*v2-y21*v1-y23*v3) + conj(v2)*(y22*v2-y21*v1-y23*v3)=2*P2_inj
12     if cGen2Reg:
13         v2*conj(v2)=V2_sp^2      // if node is PV
14     else:
15         v2*conj(y22*v2-y21*v1-y23*v3)-conj(v2)*(y22*v2-y21*v1-y23*v3)=2i*Q2_inj
16     end
17     // node 3
18     v3*conj(y33*v3-y23*v2) = S3_inj; conj(v3)*(y33*v3-y23*v2) = conj(S3_inj)
19  // End of inner Loop
20  Limits: //each limit group can process one or more limits
21     group [name="Gen2"]:
22         // PV generator on node 2
23         if cGen2Reg:
24             Q2_inj = imag(v2*conj(y22*v2-y21*v1-y23*v3))
25             if Q2_inj <= Q2_inj_min [signal=TooLow]:
26                 cGen2Reg = false  // disable voltage regulation on gen 2
27                 Q2_inj = Q2_inj_min  // set generator to PQ with given minQ
28             else:
29                 if Q2_inj >= Q2_inj_max [signal=TooHigh]:
30                     cGen2Reg = false  // disable voltage regulation on gen 2
31                     Q2_inj = Q2_inj_max  // set generator to PQ with maxQ
32                 end
33             end
34         end
35     end
36  // End of outer loop
37  end
```

Fig. 6: A code snippet illustrating the use of if-else statements for modeling a PV node with reactive power generation limits (Example 3).

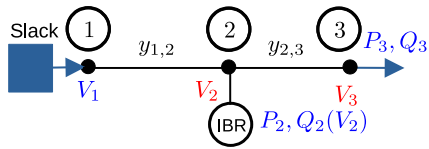

Fig. 7: A three-node network to demonstrate power flow modeling with an IBR and PQ node (the blue text indicates known quantities, while the red text represents unknown variables).

after the first group is cleared from violations, similar to the time delay controller groups introduced in [25].

*2) switch-case statement:* Fig. 7 replaces the generator at node 2 with an IBR having the Volt-VAr control characteristic shown in Fig. 8 [26], whose parameters are defined on lines 9-14 in Fig. 9. Modeling of the Volt-VAr characteristic is possible using `if-else` statements; it is, however, more straightforwardly accomplished using the `switch-case` statement on lines 19-24. The real and reactive power injection equations at nodes 2 and 3 are presented using real variables in lines 17, 18, 27, and 28. During the `NLEs` iterative solution, the

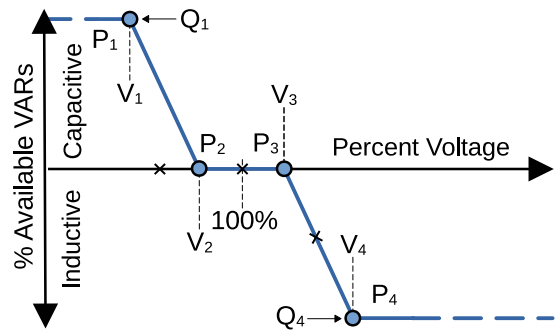

Fig. 8: IBR Volt-VAr control curve.

`switch-case` expression dictates the segment of the Volt-VAr curve that corresponds to the IBR's operating point.

*C. Iterative Simulation*

Iterative simulation is achieved through the `Repeats` group. It is suitable for continuation power flow applications

```
...
Vars [out=true]:
   ...
   Q2_inj=0    //IBR Q - initial value set to zero
Params:
   ...
   V2_sp=1.01; P2_inj=0.2  //IBR V and P
   S_ibr_rating=1.5
   cLim=0.44   //default multiplicator for max and min reactive power injection
   V_2_sp=1.02 //IBR set point
   V_dead_max = V_2_sp+0.02; V_dead_min = V_2_sp-0.02
   V_reg_max  = V_2_sp+0.08; V_reg_min  = V_2_sp-0.08
   k1 = -cLim*S_ibr_rating/(V_dead_min-V_reg_min)
   k2 = -cLim*S_ibr_rating/(V_reg_max-V_dead_max)
NLEs:
   // node 2 (IBR node with default VAr regulation)
   v_2^2*aY22*cos(θ_22) + v_2*(aY21*v_1*cos(δ_2-θ_21-δ_1) + aY23*v_3*cos(δ_2-θ_23-δ_3)) = P2_inj
   -v_2^2*aY22*sin(θ_22) + v_2*(aY21*v_1*sin(δ_2-θ_21-δ_1) + aY23*v_3*sin(δ_2-θ_23-δ_3)) =
       Q2_inj
   switch:
      case v_2 < V_reg_min -> Q2_inj=cLim*S_ibr_rating
      case v_2 < V_dead_min -> Q2_inj=cLim*S_ibr_rating+k1*(v_2-V_reg_min)
      case v_2 > V_reg_max -> Q2_inj=-cLim*S_ibr_rating
      case v_2 > V_dead_max -> Q2_inj=k2*(v_2-V_dead_max)
      default -> Q2_inj = 0
   end
   // node 3 - PQ load
   v_3^2*aY33*cos(θ_33) + v_3*(aY32*v_2*cos(δ_3-θ_32-δ_2)) = P3_inj
   -v_3^2*aY33*sin(θ_33) + v_3*(aY32*v_2*sin(δ_3-θ_32-δ_2)) = Q3_inj
end
```

Fig. 9: A code snippet illustrating the use of the switch statement for modeling IBR Volt-VAr control (Example 4).

```
...
Model [type=NL reInit=true domain=cmplx name="
    PF with PV Regulating Generator and its
    limits"]:
...
//Start of the Repeat loop
ReInit:          //Repeats is used (reInit=true)
   cGen2Reg=true
   S2_inj=0.2+0i
   S1 = 0

// Start of the outer loop
// Start of the inner Loop
NLEs:
   ... //same as Example 3

// End of the inner Loop
Limits: //each limit group can process one or
    more limits
   ... //same as Example 3
// End of the outer loop

Repeats:        //repeat until fails to solve
   S3_inj -= 0.02+0.01i //increase load at
       node 3
   repeat

//End of the Repeat loop
end       //Model end
```

Fig. 10: A code snippet illustrating the use of repetitions for computing the P–V curve (Example 5).

that involve repeated executions, such as calculating the system loadability curve or the total transfer capacity. Fig. 10 is a code snippet for calculating the P-V curve for the network in Fig. 5, with the same Vars, Params, NLEs, and Limits as in the code snippet of Fig. 6. The Repeats group defined on line 20 in Fig. 10 iteratively increases the load at node three by 0.02+0.01i, as specified on line 21. The resulting P-V curves at nodes 2 and 3 are depicted in Fig. 11. The attributes of the model on line 2 include reInit=true, meaning that the parameters in the ReInit group (line 5) are reinitialized after each repetition. This results in the power flow alternating between the high and low voltage solutions shown by the zigzagged line (Fig. 11-a). However, by setting reInit=false, the most recently calculated power flow solution is used to reinitialize the following power flow computation. This results in a smooth curve (Fig. 11-b).

## IV. STATE ESTIMATION MODELING

Similar to the case of power flow, the modelSover can also be utililized for state estimation in both real [27] and complex variables [28], [29]. The model for the solution type is WLS, as the solution is based on the Weighted Least Squares normal equations approach [27]. When the type=WLS, the corresponding equations group is WLSEs, rather than NLEs, which are used for type=NL. One could also include ECs for modeling zero injection measurements [29], instead of modeling them as virtual measurements with large weights in the normal equations approach.

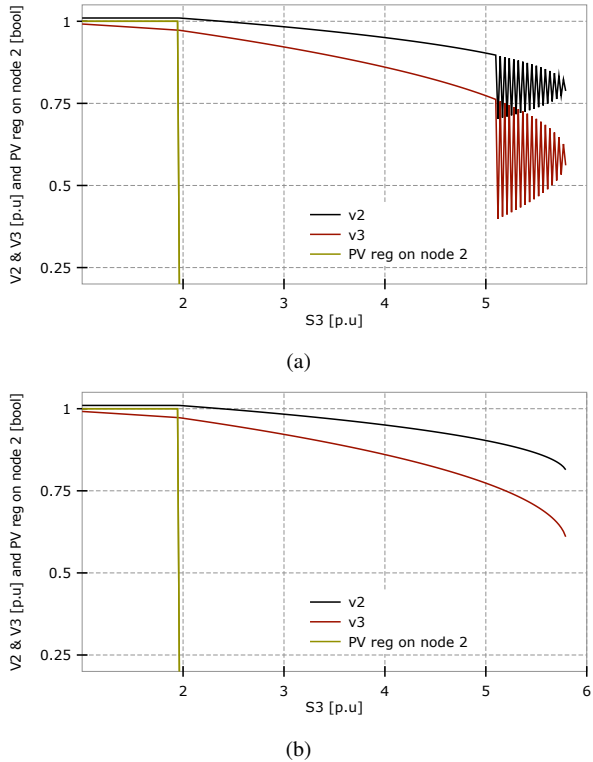

Fig. 11: Loadability characteristics at nodes 2 and 3. The zigzagged line (a) is without reinitialization, and the smooth curve (b) represents the case with reinitialization.

The code in Fig. 12 illustrates state estimation modeling for the three-node network in Fig. 2. The `Model` attribute `domain=cmplx` infers that the calculation is carried out using complex variables. The `Params` group includes the measured voltage magnitudes at all nodes (`v1_meas`, `v2_meas`, `v3_meas`), the measured complex power injection at node 3 (`S3_meas`), and the measurement weights (`w_inj` and `w_v`). The weights are mapped to the corresponding measurements in the `WLSEs` group: the phasor voltage at node 1 (lines 35-36), the complex power injection at node 3 (lines 39-40), and the voltage magnitude at nodes 2 and 3 (lines 43-44). The `ECs` group has the zero injection measurement equations at node 2 (lines 48-49). As in the case of the power flow, the solution is obtained by running the same command in the terminal: `./modelSolver PathToModFile`.

The previous example assumes that the measurements are directly included as part of the parameters. The code snippet in Fig. 13 demonstrates a `SubModel` group that is utilized to solve the power flow and obtain the true measurements. The actual measurements are then simulated by introducing Gaussian-distributed noise (lines 27-30), with the mean and standard deviation specified in lines 18-19. In the `SubModel` attributes, `copyPars=8` indicates that the first eight parameters from the `Params` list are used in the `SubModel` and therefore do not need to be redefined. As shown in

```
//The header provides solution and reporting
    details
// - It is mandatory for all models.
// - It may contain an arbitrary number of
    lines.
// - It must be terminated with the 'end'
    keyword.
Header:
  maxIter=200
  maxReps=1000
  //The 'report' attribute accepts the
      following values:
  // - Solved: Displays only the final solved
      solution.
  // - All: Shows both solved and non-solved
      attempts, including iterations.
  // - AllDetails: Includes all of the above,
      plus additional debug information
  report=AllDetails
end   //end of header
Model [type=WLS domain=cmplx name="SE_EC"]:
Vars [out=true]:      // conj="true" is by
   default
  v1=v1_ph_meas; v2=v1_ph_meas; v3=v1_ph_meas
Params:
  phiSlack=pi/4 [type=real]  // angle of
      slack

  v1_ph_meas=v1_meas*e^(phiSlack*1i)
  w_inj=10 [type=real]; w_v=1 [type=real]
  z12=0.005+0.03i; z23=0.005+0.03i;
  y21=1/z12; y23=1/z23; y22=y21+y23; y33=y23

  //measurements
  v1_meas=1   [type=real]    // real voltage
      magnitude (slack)
  S3_meas=-1-0.3i
  v2_meas=0.984267; v3_meas=0.969386

WLSEs: //Measurement equations (with
   weightings)
  //weighting factors are transfered
  //as attributes (w) using brackets

  // node 1 slack(node) Fix angle
  [w=w_v] v1 = v1_ph_meas
  [w=w_v] conj(v1) = conj(v1_ph_meas)

  // node 3
  [w=w_inj] v3*conj(y33*v3-y23*v2) = S3_meas
  [w=w_inj] conj(v3)*(y33*v3-y23*v2)=conj(
      S3_meas)

  // voltage magnitudes
  [w=w_v] v2*conj(v2) = v2_meas^2
  [w=w_v] v3*conj(v3) = v3_meas^2

ECs: //Equality constraints (no weightings)
  // node 2 (ZI: sum of currents = 0)
  y22*v2-y21*v1-y23*v3=0
  conj(y22*v2-y21*v1-y23*v3)=0
end
```

Fig. 12: A code that demonstrates state estimation modeling with equality constraints using complex variables (Example 6).

```
... 
Params:
   v1_sl=1.01*e^(1i*pi/4)         //slack node
   z12=0.005+0.03i; z23=0.005+0.03i
   y21=1/z12; y23=1/z23; y22=y21+y23; y33=y23
   S3_inj=-1-0.3i [out=true]
   v1_meas [type=real out=true]; v2_meas [type=real out=true]; v3_meas [type=real out=true]
   S3_meas [out=true]
   v1_ph_meas=v1_meas*e^(1i*pi/4)
   w_inj=10  [type=real]; w_v=1 [type=real]

SubModel [type=NL alwaysOn=true eps=1e-4 maxIter=200 domain=cmplx copyPars=8 name="Prepare
    inputs for the main probelm with RND"]:
   Vars [conj=true out=true]:
      v2=v1_sl; v3=v1_sl
   Params:
      v1=v1_sl
   Distribs:
      g1 [type=Gauss mean=0 dev=0.02]
      g2 [type=Gauss mean=0 dev=0.03]
   NLEs:
      // node 2
      y22*v2-y21*v1-y23*v3=0; conj(y22*v2-y21*v1-y23*v3)=0
      // node 3
      v3*conj(y33*v3-y23*v2) = S3_inj; conj(v3)*(y33*v3-y23*v2) = conj(S3_inj)
   PostProc:
      // transfer initial values to parent problem with noise
      @main.v1_meas.real = abs(v1) + real(rnd(g1))
      @main.v2_meas.real = abs(v2) + real(rnd(g1))
      @main.v3_meas.real = abs(v3) + real(rnd(g1))
      @main.S3_meas = v3*conj(y33*v3-y23*v2) + rnd(g2)
end       //end of submodel
WLSEs:
   ... // same as Example 6
ECs:
   ... // same as Example 6
end
```

Fig. 13: A code snippet illustrating the use of a `SubModel` for calculating measurements aided by a power flow (Example 7).

the framework in Fig. 1, the solution of the `SubModel` is followed by the state estimation calculation as defined by `WLSEs` and `ECs`.

## V. MATPOWER CONVERTERS AND COMPARISONS

Given the popularity of MATPOWER [2] for power flow-based applications, the GitHub platform [23] includes a Python script to convert MATPOWER case files into the `modelSolver` input file format, controlled via an XML config.xml file. It requires Python 3.7 or higher for full compatibility. The script includes sections for options, variables, and limits. The options settings allow selecting the generation of the mathematical representation of voltages and admittances using polar, rectangular, or complex format. Variable names can be specified as either plain alphanumeric or using Greek/special symbols. The script also defines power categories and ZIP (impedance-current-power) load models. To run the script, the user types `python3 matp2modl.py caseX.m` from the terminal.

The validity of the converter output was established by comparison with the power flow results from MATPOWER [2]. Five test systems (case5, case9, case30, case118, and case300) were converted from the MATPOWER input to `modelSolver`'s format. The conversion used settings that specified expressions in polar, rectangular, and complex variables, resulting in a total of 15 generated cases. In each instance, the power flow results from `modelSolver` corresponded to those of MATPOWER, with an absolute error less than 1E-6.fFigure f

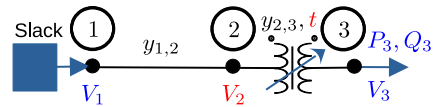

Fig. 14: A three-node network to demonstrate power flow modeling with a zero injection node, a PQ node, and an LTC transformer (the blue text indicates known quantities, while the red text represents unknown variables).

The converters enable users to generate `modelSolver` files and modify them to expand the features of the power flow solution beyond those offered by MATPOWER. For instance, consider the network in Fig. 14 that includes a Load Tap-Changing (LTC) transformer between nodes 2 and 3. MATPOWER can obtain a power flow solution with a fixed tap

```
1   ...
2   Vars [out=true]:
3   ...
4   t=1 //tap position
5   Params:
6   ...
7   cLTC23Reg=true [out=true]  //if true -> not discretized, otherwise discretized
8   deltaLTC = 0.0125        //LTC step size
9   V_3_sp = 1         //regulation set point (mid of deadband)
10  t_fix = 1          //fixed value of tap position
11  t_min = 1 - 10*deltaLTC   //min tap position
12  t_max = 1 + 10*deltaLTC   //max tap position
13  LTC_pos = 0 [type=int out=true] //tap position (relative to neutral position)
14
15  NLEs:
16      // node 2 (ZI)
17      aY21*v_1*cos(θ_21+δ_1) + v_2*aY*(1+t^2)*cos(θ_22+δ_2) + t*aY23*v_3*cos(θ_23+δ_3)=0
18      aY21*v_1*sin(θ_21+δ_1) + v_2*aY*(1+t^2)*sin(θ_22+δ_2) + t*aY23*v_3*sin(θ_23+δ_3)=0
19      // node 3
20      v_3^2*aY33*cos(θ_33) + t*v_3*(aY32*v_2*cos(δ_3-θ_32-δ_2)) = P3_inj
21      -v_3^2*aY33*sin(θ_33) + t*v_3*(aY32*v_2*sin(δ_3-θ_32-δ_2)) = Q3_inj
22      switch:
23          case cLTC23Reg -> v_3 = V_3_sp
24          default -> t=t_fix
25      end
26  IterPostP:
27      LTC_pos = round((t-1)/deltaLTC, 0)
28  Limits: //each limit group can process one or more limits
29      group [name="LTC" enabled=true]:
30          // PV generator on node 2
31          if cLTC23Reg:
32              switch:
33                  case t <= t_min [signal=TooLow] -> t_fix=t_min //min LTC limit
34                  case t >= t_max [signal=TooHigh] -> t_fix=t_max //max LTC limit
35                  default [signal=Rounding] -> t_fix=disc(t, 1, deltaLTC) //calculate discrete LTC
                        position
36              end
37              t = t_fix
38              cLTC23Reg=false     // fix LTC position (its on min, max, or rounded/discr.)
39          end
40      end   //end group
41  PostProc:
42      LTC_pos = round((t-1)/deltaLTC, 0) //calculate SCADA position (integer value)
43  end
```

Fig. 15: A code snippet illustrating the use of the switch statement for modeling an LTC transformer (Example 8).

position. However, modifying the exported `modelSolver` input file, as shown in Fig. 15, enables automatic adjustment of the tap position for voltage regulation at node 3. The Iteration Post Processing (`IterPostP` on line 26) and the Post Processing (`PostProc` on line 41) calculate the LTC position during iterations and at the converged solution, respectively.

## VI. CONCLUSION

The paper introduced `modelSolver` [23], a freely available software package for power system modeling and computing applications, explicitly focusing on load flow control and state estimation. Unlike available packages, `modelSolver` adopts an open-box approach, allowing users to specify custom models, such as those that could be associated with emerging power flow control devices and IBRs. The `modelSolver` is built on a new framework that allows users to create intuitive mathematical models without the need for traditional programming constructs. The models can be specified in the conventional rectangular and polar coordinates, and also in the more straightforward complex variable approach. It furthermore allows models that involve conditional statements, such as piecewise control curves, and iterative simulations. The software package comes with a MATPOWER converter that enables the user to import cases and further modify their model operation to scenarios beyond MATPOWER's hard-coded models. Future research aims to develop the `modelSolver`'s capabilities to handle differential-algebraic models for stability analysis, develop built-in explicit and implicit solvers, and create a graphical user interface.

**Izudin Džafić** (M'05-SM'13) received his Ph.D. degree from University of Zagreb, Croatia in 2002. He is currently a Professor in the Department of Automatic Control and Electronics, Faculty of Electrical Engineering, at the University of Sarajevo, Bosnia and Herzegovina. From 2002 to 2014, he was with Siemens AG, Nuremberg, Germany, where he held the position of the Head of the Department and Chief Product Owner (CPO) for Distribution Network Analysis (DNA) R&D. He holds 8 US and international patents. His research interests include power system modeling, development and application of fast computing to power systems simulations. Dr. Džafić is a member of the IEEE Power and Energy Society and the IEEE Computer Society.

**Rabih Jabr** (M'02-SM'09-F'16) was born in Lebanon. He received the B.E. degree in electrical engineering (with high distinction) from the American University of Beirut, Beirut, Lebanon, in 1997 and the Ph.D. degree in electrical engineering from Imperial College London, London, U.K., in 2000. Currently, he is a Professor in the Department of Electrical and Computer Engineering at the American University of Beirut. His research interests are in mathematical optimization techniques and power system analysis and computing.